\documentclass[conference,10.5pt]{IEEEtran}
\IEEEoverridecommandlockouts
\usepackage{cite}
\usepackage{amsmath,amssymb,amsfonts}
\usepackage{algorithmic}
\usepackage{graphicx}
\usepackage{color}
\usepackage{subfig}
\usepackage{stackengine}
\usepackage{comment}
\usepackage{ulem}
\usepackage{tabularx}
\usepackage{bm}
\usepackage{bbold}
\usepackage{mathtools, nccmath}
\usepackage{multirow}
\usepackage{stfloats}

\usepackage[singlespacing]{setspace} 

\usepackage{lettrine}
\usepackage[font=scriptsize]{caption}

\usepackage[colorlinks, citecolor=blue]{hyperref} 

\usepackage[flushleft]{threeparttable} 
\usepackage{booktabs}

\usepackage{textcomp}
\usepackage{xcolor}

\usepackage[linesnumbered, ruled]{algorithm2e}
\SetKwRepeat{Do}{do}{while}%

\SetCommentSty{mycommfont}

\def\BibTeX{{\rm B\kern-.05em{\sc i\kern-.025em b}\kern-.08em
		T\kern-.1667em\lower.7ex\hbox{E}\kern-.125emX}}
    
\usepackage{orcidlink}

\usepackage{xspace}
\usepackage{ifthen}
\usepackage[inline]{enumitem}

\newboolean{showcomments}
\setboolean{showcomments}{true}
\ifthenelse{\boolean{showcomments}}
{
}

\usepackage{subfig}
\usepackage{fancyhdr}

\newcommand*{\affmark}[1][*]{\textsuperscript{#1}}

\begin{document}

\title{\huge Channel Selection for Wi-Fi 7 Multi-Link Operation via Optimistic-Weighted VDN and Parallel Transfer Reinforcement Learning}

\author{\IEEEauthorblockN{ Pedro Enrique Iturria-Rivera, \IEEEmembership{Graduate Student Member,~IEEE}, Marcel~Chenier\affmark[2], Bernard~Herscovici\affmark[2],\\ Burak~Kantarci, \IEEEmembership{Senior Member,~IEEE} and Melike Erol-Kantarci, \IEEEmembership{Senior Member,~IEEE}}
\IEEEauthorblockA{\affmark[1]\textit{School of Electrical Engineering and Computer Science, University of Ottawa, Ottawa, Canada}}  \affmark[2]\textit{NetExperience., Ottawa, Canada}\\
Emails:\{pitur008, burak.kantarci, melike.erolkantarci\}@uottawa.ca,  \{marcel, bernard\}@netexperience.com
\vspace{-1em}}

\maketitle
\begin{abstract}
\textbf{Dense and unplanned IEEE 802.11 Wireless Fidelity (Wi-Fi) deployments and the continuous increase of throughput and latency stringent services for users have led to machine learning algorithms to be considered as promising techniques in the industry and the academia.  Specifically, the ongoing IEEE 802.11be EHT ---Extremely High Throughput, known as Wi-Fi 7--- amendment propose, for the first time, Multi-Link Operation (MLO). Among others, this new feature will increase the complexity of channel selection due the novel multiple interfaces proposal. In this paper, we present a Parallel Transfer Reinforcement Learning (PTRL)-based cooperative Multi-Agent Reinforcement Learning (MARL) algorithm named Parallel Transfer Reinforcement Learning Optimistic-Weighted Value Decomposition Networks (oVDN) to improve intelligent channel selection in IEEE 802.11be MLO-capable networks. Additionally, we compare the impact of different parallel transfer learning alternatives and a centralized non-transfer MARL baseline. Two PTRL methods are presented: Multi-Agent System (MAS) Joint Q-function Transfer, where the joint Q-function is transferred and MAS Best/Worst Experience Transfer where the best and worst experiences are transferred among MASs. Simulation results show that oVDN$_g$ --only the best experiences are utilized-- is the best algorithm variant. Moreover, oVDN$_g$  offers a gain up to $3\%$, $7.2\%$ and $11\%$ when compared with VDN, VDN-nonQ and non-PTRL baselines. Furthermore, oVDN$_g$ experienced a reward convergence gain in the $5$ GHz interface of $33.3\%$  over oVDN$_b$ and oVDN where only worst and both types of experiences are considered, respectively. Finally, our best PTRL alternative showed an improvement over the non-PTRL baseline in terms of speed of convergence up to 40 episodes and reward up to $135\%$. }
\end{abstract}


\section{Introduction}

\lettrine[findent=1pt]{\textbf{R}}{ }ecent IEEE 802.11 Wireless Fidelity (Wi-Fi) documentation introduces a new standard named 802.11be EHT --Extremely High Throughput-- or Wi-Fi 7, which will become the replacement for the former 802.11ax amendment. The release of the final draft is expected by May, 2024 \cite{timeline} adding to the new set of modifications a modulation rate increase up to 4096 QAM, channel bandwidth up to 320 MHz and more importantly the addition of  Multi-Link Operation (MLO) and Multiple Resource Units (MRU) capabilities \cite{Garcia-Rodriguez2021}.  These new features will allow Wi-Fi technology to adapt to the continuously increasing demand in terms of throughput and low-latency services such as Virtual Reality, Live Streaming and Ultra-High Definition video in dense deployments. 

\begin{figure}[t]
\center
  \includegraphics[scale=0.35]{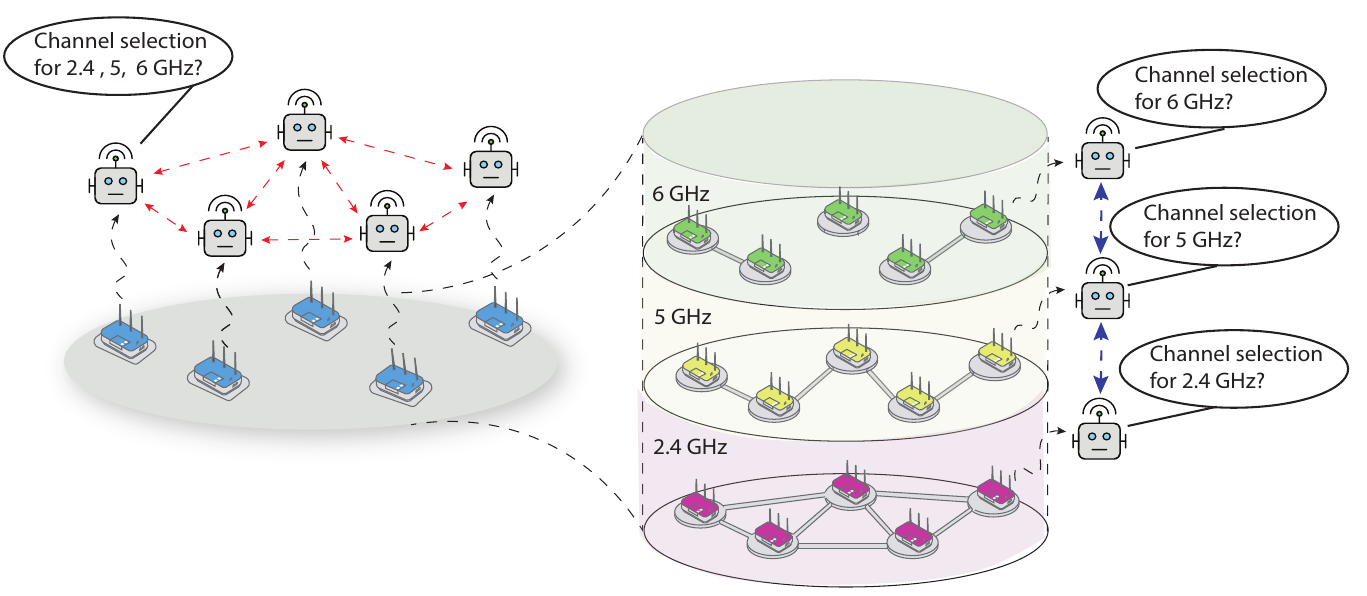}
  \setlength{\belowcaptionskip}{-5pt}
  \caption{ Instead of utilizing one Multi-Agent System (MAS) where each agent provides a joint action selection for all interfaces, in the context of MLO, we divide it in three MASs. Each MAS provides an action selection per interface and parallel transfer learning is leveraged to improve learning.} 
  \label{system_overview1}\vspace{-1em}
\end{figure} 

In this work, with the aid of Fig. \ref{system_overview1} we raise the following question: Is it beneficial to transfer knowledge in a parallel fashion among MASs? To do so, we propose dividing a complex problem in less complex ones to enable the utilization of Parallel Transfer Reinforcement Learning (PTRL) methods. Thus, we study PTRL to improve channel selection in the context of MLO in IEEE 802.11be. MLO adds an extra dimension to channel selection in IEEE 802.11 networks since it permits that Multi-Link Devices (MLD)s such as terminals and Access Points (APs) use their available interfaces for multi-link communications. This means that not only channel selection should be performed concurrently in one interface, say $2.4$ GHz;  but this procedure needs to be repeated for the rest of the interfaces. Specifically, we propose leveraging Multi-Agent Reinforcement Learning (MARL) and  PTRL to optimize channel selection in IEEE 802.11be MLO-capable networks. MARL has been brought to the attention of the wireless networks research community due its realistic approach \cite{iturria}. In addition, transfer learning has also proven its applicability and positive implications in improving learning convergence and Key Importance Metrics (KPIs) in wireless networks-based on RL multi-agent solution \cite{iturria2}. However, in this work, we utilize a less explored area of transfer learning called parallel transfer learning that proposes concurrent knowledge transfer without the source/target paradigm of the well-known sequential transfer learning. More specifically we put our interests on parallel transfer reinforcement learning among Multi-Agent Systems (MASs). To enable Machine Learning (ML)-based technology and others in Wi-Fi, a project named OpenWiFi \cite{TIP2022} propose disaggregating the Wi-Fi technology stack by utilizing open source software for the Access Point (AP) firmware operating system. Such technology enables the integration with diverse cloud-based services in the form of ML-based applications such as the one proposed in this work. 

The rest of this paper is organized as follows. Section \ref{Section2} presents the existing works related to parallel transfer learning and RL-based channel selection in Wi-Fi. In Section \ref{section3} a brief system model is described. Section \ref{section4} introduces the proposed scheme Parallel Transfer Reinforcement Learning Optimistic-Weighted VDN (referred as oVDN) and its design considerations, including the PTRL methods and the Markov Decision Process (MDP). Additionally, section \ref{section5} presents a study over the proposed scheme's performance. Finally, section \ref{Section6} concludes the paper.


\section{Related work}\label{Section2}
Channel selection algorithms in Wi-Fi have been of great interest in the recent literature. However, channel selection in IEEE 802.11be MLO has not been studied due its novelty. In addition, to the best of our knowledge, this is the first work that tackles Parallel Transfer Reinforcement Learning among MASs and applies such technique in the context of channel selection in IEEE 802.11be. Below, we list some of the works related to parallel transfer learning and channel selection.

In \cite{Taylor2013} the authors utilized parallel transfer learning among intelligent agents to coordinate electrical vehicle charging to prevent transformer overload in smart grid scenarios. This work introduces, for the first time, parallel transfer learning and discusses the advantages over the sequential transfer learning techniques. Moreover, the same authors present in a later work \cite{Taylor2019} a survey study where possible alternatives to employ parallel transfer learning is considered.
On the other hand, channel selection has been studied in the previous Wi-Fi standards where only Single Link Operation is considered. For instance, in \cite{Wang2018}, the authors study channel selection for a single user utilizing a Deep Q-Network agent. In this work, at the beginning of each time slot, the user selects a channel and receives a reward based on how successful the transmission was. Additionally, in \cite{Kishimoto2020}, a Q-Learning based algorithm is utilized in the selection of the channel/subframe in LTE-Wi-Fi scenarios. In the previous work, the authors measure the effectiveness of the algorithm based on the channel utilization and channel utilization fairness among the APs. In the next section, we present the description of the system model utilized in this work.

\section {System Model} \label{section3}

In this work, we utilize an IEEE 802.11be network with a predefined set of $\mathcal{N}$ APs and $\mathcal{P}$ stations attached per AP. All APs are MLO-capable with a number of available interfaces $n_f$.  All APs and stations support up to a maximum of 16-SU multiple-input multiple-output (MIMO) spatial streams. The path loss model corresponds to the enterprise model described in \cite{IEEEP802.112015}. In addition, we consider an adaptive control rate based on Signal-to-Interference-Noise Ratio (SINR) with a maximum 4096 QAM modulation rate.

\section{Parallel Transfer Reinforcement Learning Optimistic Weighted VDN (oVDN) } \label{section4}
In this section, we discuss the details and considerations taken in the design of the Parallel Transfer Reinforcement Optimistic Weighted VDN (oVDN) architecture. 

\subsection{Value Decomposition Networks (VDN)}\label{vdn}
Several MARL algorithms can been found throughout the literature. According to \cite{Gronauer2022} MARL can be classified either as Independent Learning (IL), Centralized-Training Decentralized Execution (CTDE) or Centralized Learning (CL) depending on how action selection and training is performed among agents. In some challenges, IL has shown good performance such as in \cite{witt2020}, however instability is typically observed due the concurrent exploration/training of the agents \cite{Lowe2017}. On the other side, CL assumes complete access to the agents' information and hence, it becomes a single agent problem. However, CL approaches are not realistic and can present several issues in terms of increasing action and state space. Finally, CTDE approaches allow to have middle ground between the two previously described techniques with an independent action selection and a centralized training step. Value Decomposition Networks (VDN) is proposed in \cite{Sunehag2018} and can be classified as a fully-cooperative CTDE algorithm that lies between Independent Q-Learning and Centralized Q-Learning. VDN builds upon the assumption that the multiagent system joint action-value function can be decomposed into individual agent's value functions as:
\begin{equation}
    Q_{tot}(\{\bm{h}^1,...,\bm{h}^n\}, \{\bm{a}^1,...,\bm{a}^n\}) \approx  \sum_{n=1}^{|\mathcal{N}|} \bar{Q}^n(h^n, a^n)
    \label{math1}
\end{equation}

where $\bar{Q}$ corresponds to the $n^{th} \in \mathcal{N}$ (the agents are considered to be APs in this work) agent's policy that only depends on its individual observation history,   $\{\bm{h}^1,...,\bm{h}^n\}$ and $\{\bm{a}^1,...,\bm{a}^n\}$ corresponds to the tuple of histories and actions of the $n^{th}$ agent, respectively. 
As mentioned previously, VDN assumes full cooperation among agents which allows to define a team reward. Thus, the team reward can be decomposed as: 
\begin{align}
    r_{tot} =  \sum_{n=1}^{|\mathcal{N}|} r^n(h^n, a^n)
\end{align}
Then, from the perspective of one agent (say $n=1$) the additive joint Q-function can be defined as:
\begin{equation}
\begin{split}
    Q_{tot}(\bm{s}, \bm{a}) =  \mathbb{E}[\sum_{t=1}^{\infty} \gamma^{t-1}r_{tot}(\bm{s}_t, \bm{a}_t)|\bm{s}_1 = \bm{s}, \bm{a}_1 = \bm{a}]\\
    =  \mathbb{E}[\sum_{t=1}^{\infty} \gamma^{t-1}r^{1}(h_t^1, a_t^1)|\bm{s}_1 = \bm{s}, \bm{a}_1 = \bm{a}] + \\ \sum_{n=2}^{|\mathcal{N}|}\mathbb{E}[\sum_{t=1}^{\infty} \gamma^{t-1}r^{n}(h_t^n, a_t^n)|\bm{s}_1 = \bm{s}, \bm{a}_1 = \bm{a}]\\
    =: \tilde{Q}_{tot}^1(\bm{s}, \bm{a}) + \sum_{n=2}^{|\mathcal{N}|}\tilde{Q}_{tot}^n(\bm{s}, \bm{a}) \\ \approx  \bar{Q}_{tot}^1(h^1, a^1) + \sum_{n=2}^{|\mathcal{N}|}\bar{Q}_{tot}^n(h^n, a^n)
\end{split}
\label{math3}
\end{equation}
where $\gamma$ corresponds to the discount factor. Finally, it can be seen that the last equation on (\ref{math3}) corresponds to Equation \ref{math1}.

\begin{figure*}[t]
\center
  \includegraphics[scale=0.65]{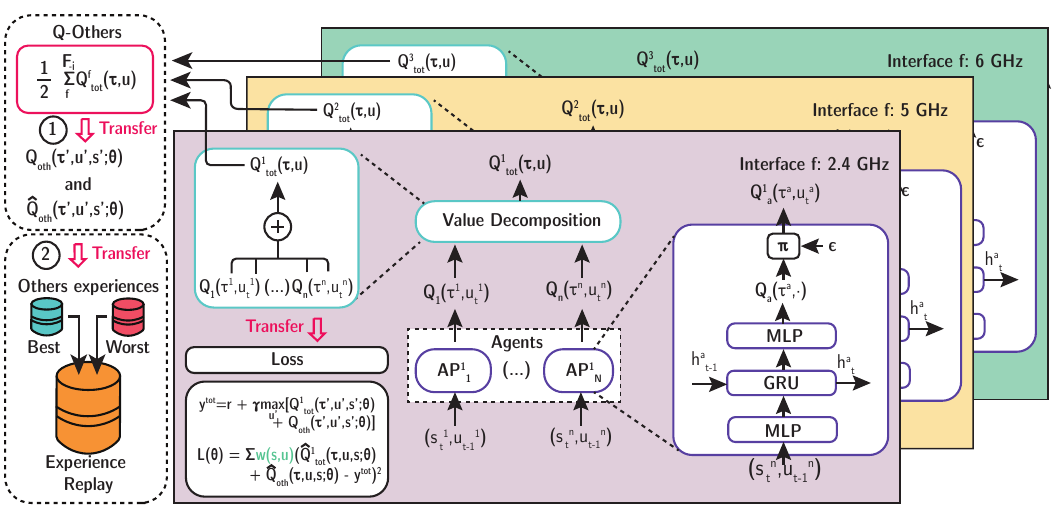}
  \setlength{\belowcaptionskip}{-5pt}
  \caption{Overview of the oVDN algorithm } 
  \label{system_overview}
  \vspace{-5mm}
\end{figure*} 

\subsection{Optimistic Weighted VDN}\label{soft-actor}

In \cite{Rashid2020}, the authors propose two weighting schemes for the QMIX algorithm \cite{Rashid2018}. The QMIX algorithm, similar to VDN is a fully cooperative algorithm that performs value function factorization among the agents. Instead of using an additive approach as mixing strategy, QMIX use hypernetworks that enforce monotonic Q-function behavior. However, QMIX assumes equal importance among actions which can lead to optimal policy failure. To solve the previous issue, the authors propose the Optimistically-Weighted QMIX that enables the assignation of weights per Q-function and consequently take better joint actions. The aforementioned issue also affects VDN due the equally weighted assumption in its additive approach. The weighting parameter can be defined as follows: 
\begin{align}
     w(o,\bm{a}) =\begin{cases} 
                            1  &  Q_{tot}( \bm{s},\bm{a}; \theta') < y_i,  \\
                               \alpha &  otherwise \\                   
                            \end{cases}
                            \label{opw}
\end{align}

where $y_i = r_i + \gamma Q_{tot}(s,a; \theta)$, $\theta$ the target policy, $\theta'$, the evaluation policy and $\alpha \in (0,1]$ a predefined weight. It can be seen that this technique downweights the contribution of those actions that are considered suboptimal.

\subsection{Multi-Agent Parallel Transfer Reinforcement Learning (MA-PTRL)}
Transfer Reinforcement Learning is a technique employed to accelerate convergence and improve learning in RL. The idea comes from human psychology where learned actions or concepts can be utilized to speed up the acquisition of new knowledge \cite{Taylor2009}. In the context of multi-agent RL, transfer learning typically happens in a sequential and unidirectional flow, from agents defined as source or teacher agents towards agents with some or minimal knowledge named target or student agents. Evidently, in some scenarios, the assumption of the existence of source-target-agent relationship is not known; in other words there is no hint which agent is the source or the target. To address the previously mentioned issue, Parallel Transfer Reinforcement Learning (PTRL) in intra-multi-agent systems is introduced in \cite{Taylor2013} where source and target agents run concurrently. Thus, knowledge transfer occurs in a parallel and bidirectional way. Consequently, it removes the need of having a set of previously trained source agents, as required by sequential RL.

In this work, to the best of our knowledge, we introduce for the first time PTRL for cooperative inter-multi-agent systems. We identify two PTRL types: Intra-PTRL that corresponds to the scenario where agents in a MAS  concurrently transfer knowledge and inter-PTRL where transfer learning occurs concurrently among MAS systems. Employing this technique can have several benefits and direct implications on the definition of the RL problem such as reduction of the action and state space as well as convergence and learning improvement. However, the MASs involved in any PTRL application must have an inner relationship to obtain the expected behavior as needed in sequential transfer learning. Moreover, a careful design must be taken in the MDP definition to avoid negative transfer. For example, we realize that scaling rewards among MAS systems do have an impact on the effectiveness of Multi-Agent Parallel Transfer Reinforcement Learning (MA-PTRL). In the next subsection, we introduce the PTRL methods used in this work.

\normalem 
\begin{algorithm}
\caption{Parallel Transfer Reinforcement Learning Optimistic-Weighted VDN (oVDN)}
\algsetup{linenosize=\tiny}
 \scriptsize
Let $\mathcal{M}$ be the set of multiagent VDN systems. For each $m^{th} \in \mathcal{M}$ VDN, initialize the agents' target and evaluation recurrent policies $\theta$ and $\theta'$, respectively. Set initial hyperparameters and set experience buffers $\mathcal{D}$ and $\mathcal{D}_{ep}$. \\
\For{environment step $t\gets1$ \textbf{to} $T$}{ 
    $\textbf{u}_{t} = \{\}$\\   
    \For{each VDN $m$ }{ 
        \For{each agent $i$ }{ 
        Observe state $s^{m,i}_t = \{a^{m,i}_{t}, i^m\}$ and append to $\tau^{m,i,a}_t=\tau^{m,i,a}_{t-1} \cup \{(s^{m,i}_t, u^{m,i}_{t-1})\} $\\
              
        $\epsilon^{m,i}_t = \epsilon^{m,i}_t - \epsilon_d$ \textbf{if} $\epsilon^{m,i}_t - \epsilon_d > \epsilon_{min}$ \textbf{else} $\epsilon_{min}$\\
        Select joint action:\\
        $u^{m,i,a}_{t} =\begin{cases} 
                        argmax_{u^{m,i,a}_{t}} [\pi^{m,i}(\tau^{m,i}_t, u^{m,i,a}_{t}; \theta^{m,'})]  & \textbf{if } \epsilon \geq \mathcal{N}(0,1),   \\
                        \text{random action from } \mathcal{I}(0,|U|^m); \\
                        |U|^m \subset \mathbb{Z}^+ &  otherwise\\
                      
         \end{cases}$\\
        
        }
        $\textbf{u}_{t} = \textbf{u}_{t} \cup \textbf{u}^{m,a}_{t}$       
    }    
    Execute $\textbf{u}_{t}$ in the environment
    \For{each VDN $m$ }{ 
        \For{each agent $i$ }{ 
            Observe next state $s^{m,i}_{t+1}$  and reward $r^{m,i}$     
        }
         $\mathcal{D}^m =  \mathcal{D}^m  \cup \{(\textbf{o}^{m}_t, \textbf{u}^{m,a}_{t}, r^m_t, \textbf{o}^{m}_{t+1})\}$ 
         $\mathcal{D}^m_{ep} =  \mathcal{D}^m_{ep}  \cup \{(\textbf{o}^{m}_t, \textbf{u}^{m,a}_{t}, r^m_t, \textbf{o}^{m}_{t+1})\}$
     }        
    \If{$t$ \textbf{mod} $n_{steps} = 0$}{

       \For{each VDN $m$ }{

                           {\fontfamily{pcr}\selectfont
\# \textcolor{red}{(1)} Transfer $Q_{tot}$ among MAS.   
}\\
            $\bm{y}_B = r_B + \gamma (\bm{Q}^m_{B,tot} + \sigma \frac{1}{2}\sum_{n}^{\mathcal{M}\setminus m} \bm{Q}^n_{B, tot}) $

            $\mathcal{L}(\theta^{m}) = \sum_{b=1}^B (\bm{w}(\bm{o}_b,\bm{u}_b)([\hat{Q}^m_{b,tot} + \sigma \frac{1}{2}\sum_{n}^{\mathcal{M}\setminus m} \hat{Q}^n_{b,tot}] - y_b)^2$
            
            $\theta^{m} \leftarrow \theta^{m,'}$\\
            \For{each VDN $n$ $\in \{\mathcal{M} \setminus m$\}}{
                    {\fontfamily{pcr}\selectfont
\# \textcolor{red}{(2)} Transfer best and worst experiences among MAS.   
}\\
                $\{(\textbf{s}^{m}_t, \textbf{u}^{m,a}_{t}, r^m_t, \textbf{s}^{m}_{t+1})\}_g = \texttt{sort}_g( \mathcal{D}^n_{ep}, n_g)$\\
                $\{(\textbf{s}^{m}_t, \textbf{u}^{m,a}_{t}, r^m_t, \textbf{s}^{m}_{t+1})\}_b = \texttt{sort}_b( \mathcal{D}^n_{ep}, n_b)$\\
                 $\mathcal{D}^m = \mathcal{D}^m \cup \{(\textbf{s}^{m}_t, \textbf{u}^{m,a}_{t}, r^m_t, \textbf{s}^{m}_{t+1})\}_g \cup \{(\textbf{s}^{m}_t, \textbf{u}^{m,a}_{t}, r^m_t, \textbf{s}^{m}_{t+1})\}_b$\\
           }
           \If{$\mathcal{D}^m > B_s$}{
            Randomly sample a batch of transitions $\bm{B}$ with size $B_s$ from $\mathcal{D}^m$\\
             \For{each batch $b \in \bm{B}$ }{
             $\bm{Q}^m_{\{1..|\mathcal{M}|\}_b} = \pi^{m}(\tau^{m}_b, u^{m}_{b}; \theta^{m,'})$\\
              $Q^m_{b,tot} = \mu( \bm{Q}^m_{\{1..|\mathcal{M}|\}_b} ) $  \\
               $\bm{\hat{Q}}^m_{\{1..|\mathcal{M}|\}_b} = \pi^{m}(\tau^{m}_b, u^{m}_{b}; \theta^{m})$\\
              $\hat{Q}^m_{b,tot} = \mu( \bm{\hat{Q}}^m_{\{1..|\mathcal{M}|\}_b}) $  
             }
            \textbf{Reset} all $D_{ep}$ buffers          
           }
        }
    }
}\medskip
 \label{ptlvdn}
\end{algorithm}

\subsection{Parallel Transfer Learning (PTRL) Methods}\label{methods}

PTRL is a challenging technique due its inherent online characteristic. Despite the benefits previously mentioned, the lack of source-target task relationship enforce a careful transfer design. Differently from other works, we utilize PTRL to improve convergence and alleviate the action/state space large dimensionality that a centralized multi-agent RL problem definition could suffer.

Let's consider the existence of $\mathcal{M}$ MASs that will concurrently transfer knowledge among them. As observed in Figure \ref{system_overview} and Algorithm \ref{ptlvdn}, we utilize two PTRL techniques in the context of cooperative MARL: \textbf{(1)} MAS Joint Q-function Transfer and \textbf{(2)} MAS Best/Worst Experience Transfer. Each of the previously mentioned techniques can be summarized as follows:

\textbf{MAS Joint Q-function Transfer}: As the name indicates the average joint Q-functions ($Q_{tot}$) of the set $\{\mathcal{M}\setminus m\}$ are used in the training stage of each individual MAS. Thus, the loss function for the $m^{th}$ MAS can be written as follows: 
\begin{equation}
     y = r + \gamma (Q^m_{tot} + \sigma \frac{1}{|\mathcal{M}\setminus m|}\sum_{i}^{\mathcal{M}\setminus m} Q^i_{tot}) 
     \end{equation}

 \begin{equation}              
            \mathcal{L} = w(o,u)([\hat{Q}^m_{tot} + \sigma \frac{1}{|\mathcal{M}\setminus m|}\sum_{i}^{\mathcal{M}\setminus m} \hat{Q}^i_{tot}] - y)^2
\end{equation}

where $\sigma$ is a tunable transfer weight, $Q_{tot}$ and $\hat{Q}_{tot}$ are target and evaluation Q-functions, respectively. Transferring the Q-function values can help to improve convergence since it suggests a consensus on the set of actions that are being selected by the MASs comprising the PTRL scheme. 

\textbf{ MAS Best/Worst Experience Transfer}: This approach relies on the nature of offline RL algorithms that utilize an experience replay as part of their policy training. Intuitively, best and worst experiences contribute to the final agent's action selection policy due the fact that the best action will receive higher rewards than worst ones. Specifically, we gather during each episode, for each $m^{th}$ MAS, $n_e$ number of sample experiences and store them in a temporal buffer $\mathcal{D}_{ep}$. As indicated in Equations (\ref{good})-(\ref{bad}), we sort $(n_g, n_b) < n_e$ samples ranking the best and worst experiences based on the ones with higher and lower rewards, respectively.  

\begin{equation}\label{good}
        \{(\textbf{s}^{m}_t, \textbf{u}^{m}_{t}, r^m_t, \textbf{s}^{m}_{t+1})\}_g = \texttt{sort}_g( \mathcal{D}^m_{ep}, n_g)
\end{equation}
\begin{equation} \label{bad}
                \{(\textbf{s}^{m}_t, \textbf{u}^{m}_{t}, r^m_t, \textbf{s}^{m}_{t+1})\}_b = \texttt{sort}_b( \mathcal{D}^m_{ep}, n_b)
\end{equation}
\begin{equation} \label{append}       
                \mathcal{D}^{\mathcal{M}\setminus m} = \mathcal{D}^{\mathcal{M}\setminus m} \cup \{(\textbf{s}^{m}_t, \textbf{u}^{m,a}_{t}, r^m_t, \textbf{s}^{m}_{t+1})\}_g \cup \{(\textbf{s}^{m}_t, \textbf{u}^{m,a}_{t}, r^m_t, \textbf{s}^{s}_{t+1})\}_b
\end{equation}

where $\textbf{s}_t^m$ , $\textbf{s}_{t+1}^m$, $r_t^m$ and $\textbf{u}_t^m$  corresponds to the state, next state, team reward and joint action, respectively. Finally, we append both experiences to the global buffer as indicated by Equation (\ref{append}).

\subsection{State space selection}
The state space is defined as follows: 

\begin{equation}
    s_{(t,f)}^n = \{a_{(t-1),f}^n, h_{(t-2),f}^n, AP_{id}^n\}
\end{equation}

where $a_{(t-1),f}^n$ corresponds to the last action taken, $h_{(t-2),f}^n$ the history action taken before it  and the $AP_{id}$ corresponds to the ID of $n^{th}$ AP. The first two terms help to alleviate partial observability via the utilization of the Gated Recurrent Unit (GRU) layer in the Q-function as observed in Figure \ref{system_overview}. The previous definition refers to the action space of a single agent in any of the MASs of our algorithm.

\begin{figure}[t]
\center
  \includegraphics[scale=0.58]{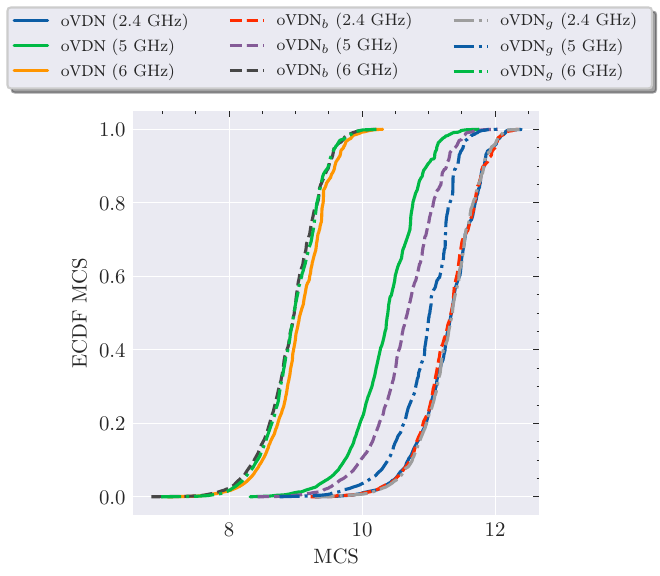}
  \setlength{\belowcaptionskip}{-1pt}
  \caption{MCS ECDFs comparison per interface frequency of three variants of transfer: oVDN, oVDN$_b$ and oVDN$_g$, indicating transfer best and worst, only best and only worst experiences, respectively.} 
  \label{ecdf-alt}
  \vspace{-5mm}
\end{figure}

\begin{figure}[t]
\center
  \includegraphics[scale=0.58]{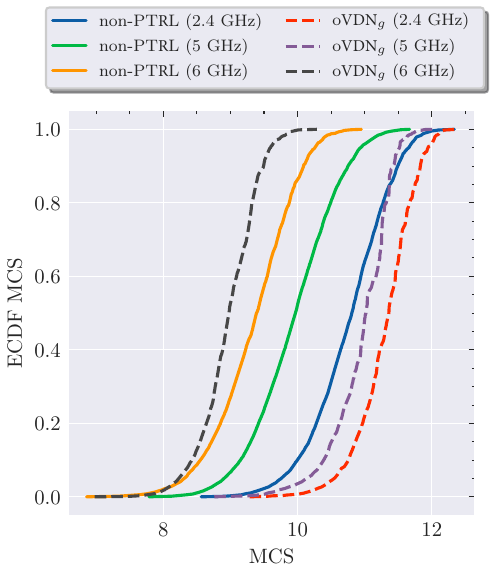}
  \setlength{\belowcaptionskip}{-1pt}
  \caption{MCS ECDFs comparison per interface frequency of the non-PTRL and our proposal using best experience transfer.} 
  \label{ecdf-non-ptl}
\end{figure}

\subsection{Action space selection}\label{action_space}
In this subsection, we present the proposed general action space: 

\begin{equation}
    a_{(t,f)}^n = \{1, ... , c_{(max,f)}\} 
\end{equation}

where $c_{(max,f)}$ corresponds to the maximum number of channels per interface frequency, and  $f \in \{2.4, 5, 6\}$ GHz indicates the frequency of each of the MASs utilized in this work.  As specified before, the current definition indicates the action space of a single agent of any of the defined MASs. Finally, the action space's size corresponds to $|a_{t,f}|!$.

\subsection{Reward function }
The general reward function for the $m^{th}$ MAS can be defined as follows:

\begin{equation}
 r_{t}^m = min(\texttt{mcs}_{f}^1, ... , \texttt{mcs}_{f}^{|\mathcal{N}|}) 
 \end{equation}

 where $\texttt{mcs} \in \{0,1, ... ,13\}$ indicates the corresponding mapped Modulation Code Selection (MCS) index given the measured SINR. The SINR comprises the effect of all APs and stations considered during the simulations. The mentioned mapping is done according the IEEE 802.11be that employs a similar MCS-SINR mapping as IEEE 802.11ax. In addition we scale the reward to $[-1,1]$.

\begin{figure*}[t]
\centering

  \includegraphics[scale=0.48]{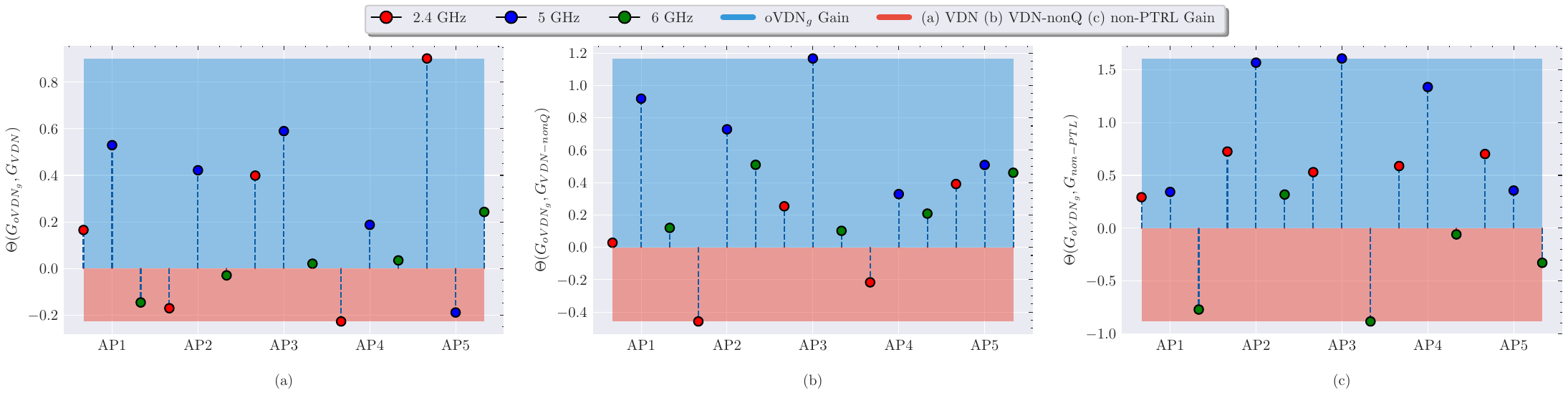}
  \caption{Gain $\Theta$ comparison per AP and interface frequency \textbf{(a)} oVDN$_g$ and VDN, \textbf{(b)} oVDN$_g$ and VDN-nonQ, \textbf{(c)} oVDN$_g$ and non-PTRL. The blue and red area indicates what technique offers the best performance. }
 \label{gain}
\end{figure*}

 A special consideration is taken in the context of channel selection in IEEE 802.11be for the MA-PTRL proposed scheme regarding the definition of the rewards.  Evidently, Q-functions depend on the rewards acquired during the interaction of the environment and as discussed in subsection \ref{methods}, Q-functions are used as part of the information transferred. Thus, different reward's scales can provoke that some of the MASs comprising the MA-PTRL algorithm can impose their action selection policy.  Specifically, we realize that MCS selection according the SINR ranges varies depending the interface frequency and channel bandwidth. Thus, we scale the rewards received by each MASs. Notice that this consideration must only be taken when there is a scaling difference between MASs' rewards in any PTRL use case. 
 
\section{Performance evaluation}\label{section5}

Simulations are performed using the flow-level simulator Neko 802.11be\cite{Lopez-Raventos2021}. The bidirectional communication between the environment simulated in Neko and the Pytorch-based RL scheme is performed via the ZMQ broker library.

\subsection{Simulation Settings}
RL parameters and simulation settings are described in Table \ref{learning_settings} and \ref{net_settings}, respectively. Furthermore, we consider the scenario where there are $|\mathcal{N}| \sim \mathcal{U}(25,40)$ users attached to each AP and the AP's locations are unknown, thus solely relying on RF feedback from the environment. In addition, $80\%$ of the users are positioned randomly in a radius $r\sim[1-8] $ m and $20\%$ within a radius of $r\sim[1-3]$ m. In the next subsection, we will discuss the performance results of the proposed scheme.  

\begin{table}
\caption{Reinforcement Learning Settings.}
\tiny
\begin{center}

\begin{tabular}{c c} 
\hline
\textbf{Parameter}&\textbf{Value} \\
\hline

Training steps & {$1.5\mathrm{e}{4}$} \\
Initial $\epsilon$ & {1} \\
$\epsilon$ decay steps & {$500$ steps} \\
$\epsilon$ minimum & {$5\mathrm{e}{-2}$} \\
Buffer $(\mathcal{D})$ size & {$2\mathrm{e}{3}$} \\
Batch size & {$64$} \\ 
Learning rate & {$8\mathrm{e}{3}$} \\
Discount factor $(\gamma)$ & {$0.99$} \\
Recurrent Layer hidden dimension & {$64$} \\
MultiLayer Perceptron hidden dimension & {$64$} \\
Weight initializer & { Orthogonal } \\
Deep Q-Network structure & {Double Q-networks} \\
Optimistic Weight $\alpha$ & {$0.1$} \\
\hline
Number of MASs & {$3, (2.4, 5, 6)$ GHz}\\
Reduced buffer $(\mathcal{D}_{ep})$ size & {$50$ }\\
Best/Worst transferred experiences & { $20\%|\mathcal{D}_{ep}|$ } \\
Transfer weight $(\sigma)$ & {1}\\
\hline
\end{tabular}

\label{learning_settings}
\end{center}
\vspace{-4mm}
\end{table}

\begin{table}
\caption{Network Settings}
\begin{center}
\resizebox{\columnwidth}{!}{%
\begin{tabular}{c c} 
\hline
\textbf{Parameter}&\textbf{Value} \\
\hline
IEEE protocol & { 802.11be } \\
Channel Bandwidth & { $20$ MHz/$40$ MHz/$80$ MHz/$160$ MHz } \\
Carrier Frequency ($f_c$) & { $2.437$ GHz/$5.230$ GHz/$6.295$ GHz } \\
Max Modulation/Max Modulation Coding Scheme & { 4096 QAM/MCS 13}\\
Number of APs per MAS $(\mathcal{N})$ & { 5 } \\
Number of Stations per AP & {  $N_A \sim \mathcal{U}(25,40)$}\\
Max Spatial Streams & { 16 } \\ 
Propagation Loss Model & { $P_{l}(d) = 40.05 + 20\text{log}(f_{c}/2.4) + 20\text{log}(\text{min}(d,10)) +$} \\ {} & {$(d>10) * 35\text{log}(d/10) + 7W$} \\
AP/STA Noise Figure & {$7$ dB}\\
AP/STA Transmission Power & {$20$/$15$ dBm}\\
CCA threshold & {-82 dBm}\\
Packet Error Rate & {$10$\%}\\

\hline
\end{tabular}
}
\label{net_settings}

\end{center}
\vspace{-7mm}
\end{table}

\subsection{Simulation Results}
We present the performance results of our proposed scheme in terms of MCS ECDF, ECDF gain and convergence,  among the baselines and the best PTRL variant.  Figure \ref{ecdf-alt} shows the MCS ECDFs of three difference variants based on what type of experiences are being transferred. The results show that the ECDF of the scenario, where only the best experiences are transferred (oVDN$_g$), contributes more positively to the channel selection and thus, higher MCS is obtained. Interestingly, only transferring worst examples offers a better performance over transferring both types of experiences. The reason behind this behavior could be explained by the fact that we utilize a fix number of worst and bad experiences and a weighted approach could be the best alternative to use.   
Additionally, in Fig. \ref{ecdf-non-ptl}, we present the results in terms of  MCS ECDFs for the centralized baseline and the best transfer variant (oVDN$_g$). It can be seen that our parallel transfer solution offers an important improvement in terms of MCS over the non-parallel baseline in two interfaces (2.4 and 5 GHz) with MCS in the range of $(10-13)$ over $(8-11)$ and a slight degradation in the 6 GHz interface. 

\begin{figure*}
\center
  \includegraphics[scale=0.50]{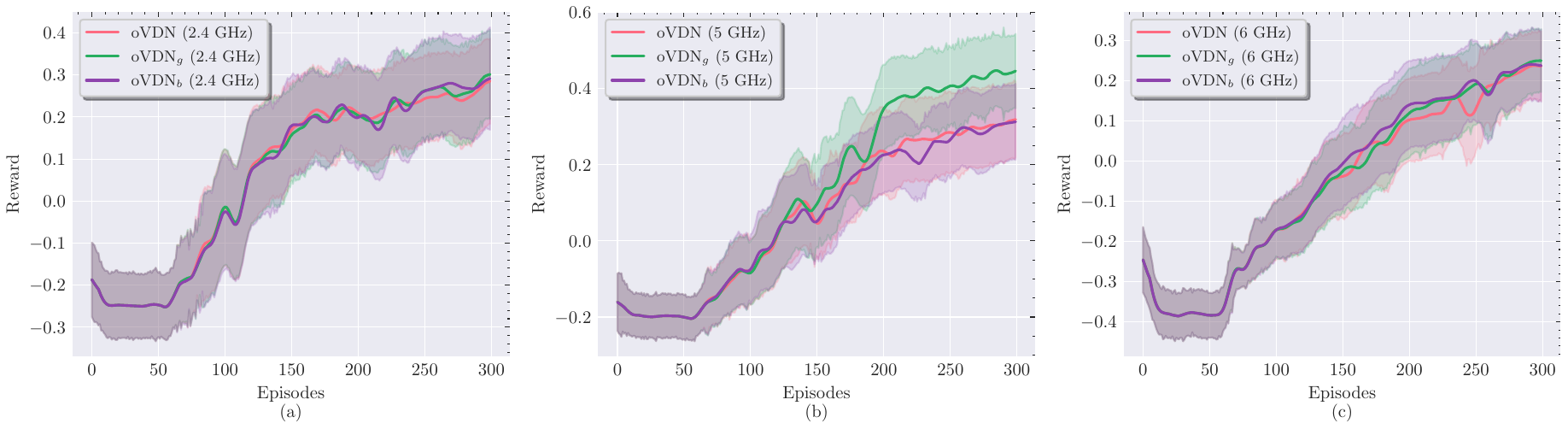}
  \caption{Reward convergence for oVDN, oVDN$_g$ and oVDN$_b$ \textbf{(a)}  $2.4$ GHz, \textbf{(b)}  $5$ GHz and \textbf{(b)}  $6$ GHz  }
 \label{convergence_between_ovdn_alternatives}
\end{figure*} 
\begin{figure*}
\center
  \includegraphics[scale=0.50]{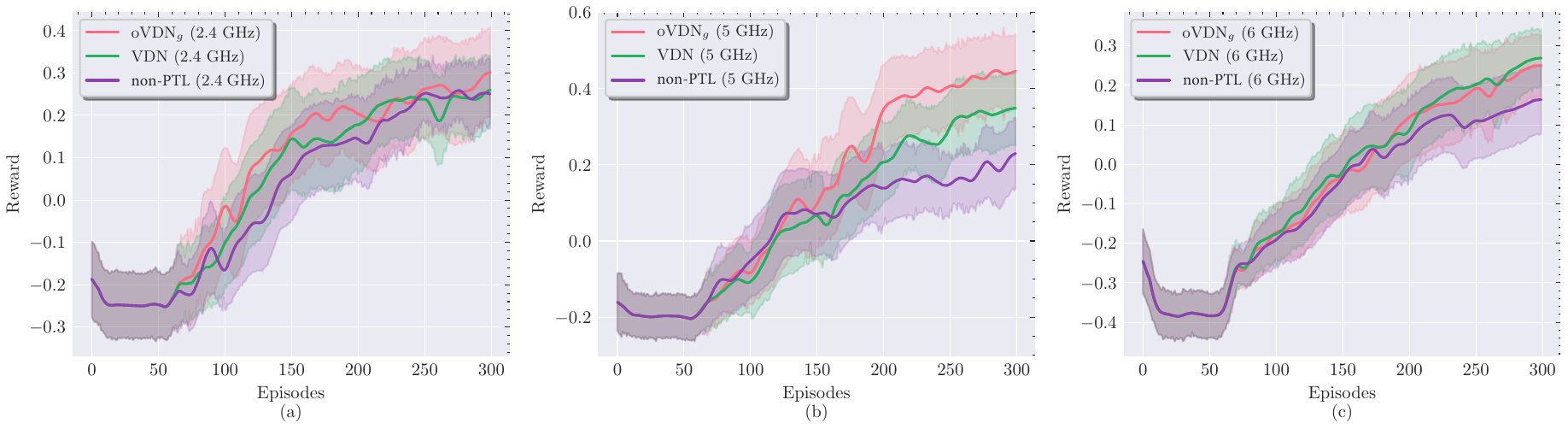}
 \caption{Reward convergence for oVDN$_g$, VDN and oVDN$_b$ \textbf{(a)}  $2.4$ GHz, \textbf{(b)}  $5$ GHz and \textbf{(b)}  $6$ GHz  }
 \label{convergence_between_vdn}
 \vspace{-5mm}
\end{figure*}

In Fig. \ref{gain}, we illustrate the obtained gain per AP and interface frequency of oVDN$_g$ over three baselines: VDN, VDN-nonQ and non-PTRL, indicating both type of experience transfer, absence of Joint Q-function transfer and the non-PTRL case, respectively.  The gain $\Theta(G_X, G_Y)$ corresponds to the area difference among ECDFs as follows:
\begin{equation}
    \begin{split}
    \Theta(G_X, G_Y) &= \int\{G_Y(x) - G_X(x)\}dx \\
    &= \int\{[1-G_Y(x)] - [1-G_X(x)]\}dx  \\
    &= \mu_X - \mu_Y 
\end{split}
\end{equation}

where $G_Y$ and $G_X$ correspond to the ECDFs of the compared variants and $\mu_X$, $\mu_Y$ are their means. The upper blue colored area indicates where the oVDN$_g$ performs better than the baselines and the red area otherwise. Consequently, we observe that the oVDN$_g$ offers a gain up to $3\%$, $7.2\%$ and $11\%$ when compared with (a) VDN, (b) VDN-nonQ  and (c) non-PTRL baselines, respectively.

Additionally, we present in Fig. \ref{convergence_between_ovdn_alternatives} a comparison in terms of convergence per frequency interface among oVDN's PTRL variants. In (a) and (c) the behavior is similar among the studied techniques, however in (b) where we show the reward convergence of the 5GHz interface, we observe an improvement of $33.3\%$ of oVDN$_g$ over oVDN$_b$ and oVDN, respectively. Finally, in Fig. \ref{convergence_between_vdn}, we can see the convergence behavior of our best performer PTRL algorithm oVDN$_g$, VDN and a non-PTRL alternative. Evidently, our proposal outperforms the non-PTRL baseline in terms of speed of convergence up to 40 episodes and reward up to $135\%$. The previous results can be explained due the non-PTRL or centralized variant's action space size. For instance, as discussed in subsection \ref{action_space}, the action space size of each MASs in the PTRL algorithm corresponds to $|a_f|^N, f \in \{2.4, 5, 6\}$, meanwhile the centralized to $|a_{2.4} \times a_{5} \times a_{6}|^N$, where $a$ indicates the set of possible actions, the $\times$ operator corresponds to the Cartesian product and $N$ the number of agents. Thus, if $|a_f|=3$, PTRL offers a reduction of action space in the order of $9^N$ which is quite considerable, especially when $N$ increases.  

To sum up, we can foresee the advantages that can potentially offer a PTRL approach among MASs such as oVDN$_g$. As observed in our presented results, faster convergence and better action selection are some of the most immediate benefits. When decomposition of a MAS problem is feasible, we advise to utilize PTRL to accelerate and improve RL performance since typically MARL suffers from slow convergence and non-optimal action selection.

\section{Conclusions} \label{Section6}
In this paper, we presented a Parallel Transfer Reinforcement Learning (PTRL) and cooperative optimistic-weighted VDN algorithm to improve radio and Reinforcement Learning (RL) related metrics such as Modulation Code Selection (MCS) and convergence speed in IEEE 802.11be channel selection. To the best to our knowledge, we studied for the first time, PTRL techniques in the context of knowledge transfer among Multi-Agent Systems (MASs). Specifically, we proposed two techniques named: \textbf{MAS Joint Q-function Transfer} and \textbf{MAS Best/Worst Experience Transfer}. The previous techniques consist in transferring the joint Q-Function of cooperative MARL algorithms such as VDN and transfer the best and worst experiences during each episode. Moreover, we presented a modification to the VDN algorithm taken from another technique named QMix that allows to weight the contribution of each agent to the joint Q-function. Furthermore, we presented the simulation results of our proposed scheme and analyze how each PTRL method affects its behavior. Results showed that the oVDN$_g$ offers an important improvement in terms of MCS over the non-parallel baseline in two interfaces ($2.4$ and $5$ GHz) with MCS in the range of $(10 13)$ over $(8 11)$. Additionally, we present a metric gain to calculate the MCS improvement. We observed that the oVDN$_g$ variant offers a gain up to $3\%$, $7.2\%$ and $11\%$ when compared with VDN, VDN-nonQ and non-PTRL baselines. In terms of convergence speed among oVDN alternatives, we show a reward convergence gain in the $5$GHz interface of $33.3\%$  over
oVDN$_b$ and oVDN. Finally, our best PTRL alternative showed an improvement over the non-PTRL baseline in terms of speed of convergence up to 40 episodes and reward up to $135\%$.  As future work, we intend to generalize PTRL to other cooperative MARL techniques.\vspace{-2mm}

\section{Acknowledgment }\label{Section7}
This research is supported by Mitacs Accelerate Program and NetExperience Inc.\vspace{-2mm}


\bibliography{biblio.bib}{}
\bibliographystyle{IEEEtran}

\end{document}